\documentclass[reprint, nofootinbib,amsmath,amssymb,aps]{revtex4-1}

\usepackage{graphicx}
\usepackage{dcolumn}
\usepackage{bm}
\usepackage{amssymb}   
\usepackage{color, colortbl}
\usepackage{textcomp}
\usepackage{ulem}
\usepackage{floatrow}
\usepackage{multirow}
\usepackage{hyperref}
\newfloatcommand{capbtabbox}{table}[][\FBwidth]

\begin{document}

\title{Phosphorene analogues: isoelectronic two-dimensional group-IV monochalcogenides with orthorhombic structure}

\author{L\'idia C. Gomes}
\affiliation{ Centre for Advanced 2D Materials and Graphene Research Centre, National University of Singapore, 6 Science Drive 2, 117546, Singapore }%
\author{A. Carvalho}
\affiliation{ Centre for Advanced 2D Materials and Graphene Research Centre, National University of Singapore, 6 Science Drive 2, 117546, Singapore }%

\date{\today}

\begin{abstract}
The group-IV monochalcogenides SnS, SnSe, GeS, and GeSe form a family within the wider group
of semiconductor `phosphorene analogues'.
Here, we used first principles calculations to investigate systematically 
their structural, electronic and optical properties, 
analyzing the changes associated with the reduction of dimensionality, from bulk to monolayer or bilayer form. 
We show that all those binary phosphorene analogues are semiconducting, with bandgap energies
covering part of the infra-red and visible range, and in most cases higher than phosphorene.
Further, we found that they have multiple valleys in the valence and conduction band, the latter with spin-orbit splitting
of the order of 19-86 meV.
\end{abstract}

\pacs{Valid PACS appear here}
\maketitle


\hyphenation{ALPGEN}
\hyphenation{EVTGEN}
\hyphenation{PYTHIA}


\section{Introduction}

Two-dimensional (2D) materials have been extensively studied ever since a monolayer graphene was 
isolated by mechanical exfoliation~\cite{geim}. 
Thereafter, the interest was promptly 
extended to other 2D materials, such as hexagonal boron nitride (h-BN), layered metal dichalcogenides (LMDCs) 
and phosphorene, to name a few~\cite{sheneve, lifeng, qing}. 
In special, phosphorene, a monolayer of black phosphorus, adopts an orthorhombic structure 
different from graphene and transition metal dichalcogenides.\cite{rodin, tomanek}
This anisotropic structure is in the origin of some of phosphorene's interesting properties, 
such as superior flexibility under tensile strain,\cite{wei-APL-104-251915}
and giant thermoelectric coefficient.\cite{fei-Nano Lett-14-6393}

However, the waved structure of phosphorene is shared by yet another class of 2D materials
that has so far eluded attention.
In the bulk form, group-IV monochalcogenides GeS, GeSe, SnS and SnSe all assume structures that can be considered derivatives 
of the orthorhombic black phosphorus~\cite{wiedemeier},
 belonging to the space group Pcmn-D$^{16}_{2h}$ 
(lower than black phosphorus, which has only one element and therefore belongs to Bmab-D$^{18}_{2h}$).
This is the $\alpha$ phase of SnS, also known as Herzenbergite, a naturally occurring but rare mineral.\footnote{Some of these 
compounds (at least SnS and SnSe) have a more symmetric $\beta$-phase, with space symmetry Cmcm-D$^{17}_{2h}$, at higher temperature.\cite{ettema}
However, according to our calculations, all of the four compounds are most stable in the $\alpha$-phase on the monolayer form.}

Currently the most important prospective application of $\alpha$-SnS is as
 absorber material for film photovoltaic (PV) cells.
Although other chalcogenide materials as CdTe 
and CuInGaSe$_2$ also show high PV efficiencies\cite{green,repins,tritsaris}, 
many factors make their usage difficult such as the high cost and toxicity of Cd~\cite{tritsaris,vidal,brad}. 
In contrast, SnS is made of abundant and nontoxic elements, 
and its optical band gap of $\sim$ 1.3~eV\cite{tritsaris, ktr} 
is right in the range of the optimal values for solar cells (1.1 to 1.5 eV).
Moreover, according to a recent study, solar conversion efficiencies achieved so far for SnS can be well beyond the
potential limit for the material due to the poor choice of band alignment in the devices.\cite{burton-apl}
In addition, group-IV monochalcogenides may show to be superior to other 2D semiconductors
in properties where anisotropy plays an important role,
as patent in the record thermoelectric coefficient recently reported for SnSe ($ZT$=2.6 at
923~K).\cite{dong-nature-508-373}

Besides, these binary `phosphorene analogues' are expected  
to reveal distinct intrinsic properties in monolayer form, as
some of the lattice symmetry operations, including inversion, 
are only present in bulk and in even-numbered layer systems.
In this sense they are different from phosphorene,\cite{dresselhaus}
where inversion symmetry prevents spin-orbit (SO) splitting.
In contrast, as shown in the present article, group-IV monochalcogenide monolayers show a large intrinsic spin-orbit splitting
at valence and conduction band valleys. 
However, even though a few theoretical and experimental works have
reported on the  electronic and optical properties of monolayer or few layer SnS,\cite{tritsaris,bablu}
monolayer properties of this group remain poorly explored.

In this work, we use first-principles calculations to investigate electronic, structural and optical properties of the 
four aforementioned group-IV monochalcogenides MX, with M=(Sn, Ge) and X=(S, Se), in the phosphorene-like $\alpha$-phase. 
We compare the properties of monolayer and bilayer with those of bulk for each of these materials,
highlighting the differences in the electronic and optical properties.

\section{Methods}

We use first-principles calculations based on density-functional theory to obtain the electronic, structural and 
optical properties of monochalcogenides. 
We employ a first-principles approach based on Kohn-Sham density functional theory (KS-DFT)\cite{ks}, as implemented 
in the {\sc Quantum ESPRESSO} code.\cite{Giannozzi2009}. The exchange correlation energy is described by the generalized 
gradient approximation (GGA) using the PBE\cite{pbe} functional. 
Interactions between valence and core electrons are described by Troullier-Martins pseudopotentials\cite{tm}. 
The Kohn-Sham orbitals were expanded in a plane-wave basis with a cutoff energy of 70~Ry, and for the charge density, a cutoff of 
280~Ry was used. The Brillouin-zone (BZ) was sampled using a $\Gamma$-centered 1$\times$10$\times$10 grid following the scheme 
proposed by Monkhorst-Pack\cite{Monkhorst1968}. For the optical properties (dielectric constant and conductivity), a finer 
1$\times$40$\times$40 grid is employed. The calculation of the spin-orbit splitting was performed using noncolinear calculations
with fully relativistic pseudopotentials.

In addition, a hybrid functional approximation for the exchange-correlation term (HSE06)~\cite{hse06}
is employed in order to give reliable results for the gap energies, which are well known to be underestimated
when employing semilocal GGA approximations.
For the hybrid functional bandstructure calculations, we used the Vienna ab initio simulation package (VASP)~\cite{vasp1, vasp2} with the
projector-augmented wave potentials~\cite{paw}.
An energy cutoff of 40~Ry was used for the plane-wave basis set and integrations
over BZ were performed using samples of 1$\times$8$\times$8 k-points for monolayers and bilayers and 4$\times$8$\times$8 k-points
for bulk structures.

For monolayer and bilayer models, we used periodic boundary conditions along the three dimensions,
with  vacuum regions of 8 and 9 \AA{}, respectively, between adjacent images in
direction perpendicular to the layers. Convergence tests with greater vacuum spacing, guarantee that this size is enough to avoid spurious interaction between neighboring images.

The optical conductivity was calculated directly from the joint density of states ie. taking into
account only direct excitations. The real part of the dielectric function is then calculated using the Kramers-Kronig relationship.

\section{Results}

\subsection{Crystal Structure}

\begin{figure}[!htb]
  \centerline{
   \includegraphics[scale=0.45]{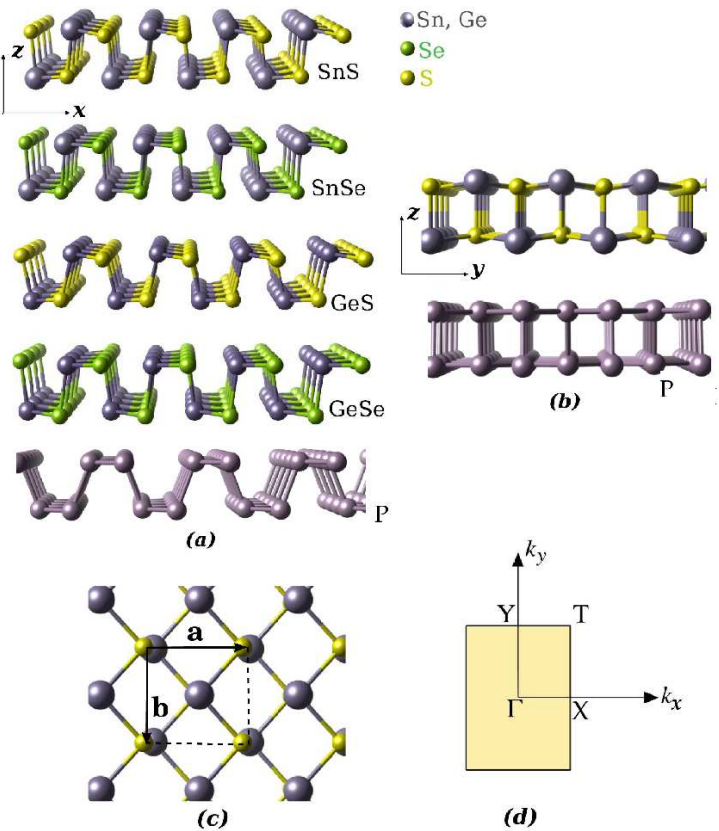}}
  \caption{\small (Color online)  
  Optimized structures of monolayers of group-IV monochalcogenides with phosphorene-like structure. 
  (a) Side view of the {\it x}-{\it z} plane for the four compounds and for phosphorene. (b) Side view of the 
  {\it y}-{\it z} plane of SnS and phosphorene. (c) Top view of the structures, with the lattice vectors 
  $\mathbf{a}$ and $\mathbf{b}$ along the {\it x} and {\it y}-directions. (d) The respective BZ and the high 
  symmetry points $\Gamma$, X, T and Y.}
  \label{unitcell-bz}
\end{figure}

The bulk $\alpha$ phase has an orthorhombic structure with eight atoms per primitive unit cell, four of each species.
The primitive unit cell contains two puckered layers, stacked on top of each other. The bilayer
is obtained by increasing the lattice supercell vector perpendicular to the plane of the layers. The monolayer has four atoms per
unit cell, as in Fig.~\ref{unitcell-bz}(c). Each atomic species is covalently bonded to three neighbors of the other atomic
species, forming zig-zag rows of alternating elements.
Thus, there is in each atom a lone pair pushing its three bonds towards a tetrahedral coordination,
just like in black phosphorus, resulting in its characteristic waved structure.

We adopt the axes system used by previous works\cite{brad, prb-41, prb-31-2180}, 
where layers are chosen to sit on the {\it x}-{\it y} plane ie. perpendicular to the {\it z}-direction. 
This is the same system conventionally used for black phosphorus, which takes the {\it y}-axis to be 
parallel to the puckering direction. Each layer has three non-trivial symmetry operations, namely a 
vertical mirror plane parallel to the xz plane, a two-fold screw rotation along an axis parallel to y, 
and a glide reflexion on a plane parallel to the xy plane.
The atomic positions, in units of the unit cell vectors $\mathbf{a}$, $\mathbf{b}$ and $\mathbf{c}$, are 
$\pm$($x$,$\frac{1}{4}$,$z$; $\frac{1}{2}+z$,$\frac{1}{4}$,$\frac{1}{2}-x$).
For bulk SnS, for example, our calculated 
fractional atomic coordinates are $x$(Sn)=$z$(Sn)=0.12, $x$(S)=0.48 and $z$(S)=0.85. 
The lattice parameters and fractional atomic positions do not deviate much from those in the parent black phosphorus structure.
As can be seen in Fig.~\ref{unitcell-bz}(a), the most noticeable difference is that the height of the atoms (along {\it z}) is no longer constant,
rather cation and anion have slightly different heights alternating along {\it x}. 
The lattice is also more compact along {\it x}, but less compact along the {\it y}-direction, to decrease the repulsion 
between the atoms of the same type aligned up along the ripples (Table~\ref{vectors}).
Full details of the structure of the  other compounds, 
including the calculated  $x$(M,X) and $z$(M,X), can be found in the Supplemental Material (SM). 

\begin{table}[ht]
\centering
\renewcommand{\arraystretch}{1.4}
\begin{tabular}{m{1.0cm} m{0.8cm} m{1.3cm} m{0.8cm} m{1.3cm} m{0.8cm} m{0.8cm} m{0.9cm}}
        \hline \hline
        & \multicolumn{2}{c}{Monolayer   } & \multicolumn{2}{c}{Bilayer     } &        \multicolumn{3}{c}{Bulk}              \\ \hline             
        &  $\mathbf{a}$  &  $\mathbf{b}$   & $\mathbf{a}$    &  $\mathbf{b}$  & $\mathbf{a}$ & $\mathbf{b}$ &  $\mathbf{c}$ \\ \hline
SnS     &     4.24       &      4.07       &    4.28         &     4.05       &   4.35       &    4.02      &   11.37       \\
SnSe    &     4.36       &      4.30       &    4.42         &     4.25       &   4.47       &    4.22      &   11.81       \\
GeS     &     4.40       &      3.68       &    4.42         &     3.67       &   4.40       &    3.68      &   10.81       \\
GeSe    &     4.26       &      3.99       &    4.31         &     3.97       &   4.45       &    3.91      &   11.31       \\ \hline
P       &     4.60       &      3.30       &    4.57         &     3.31       &   4.57       &    3.51      &   11.69       \\ \hline \hline
\end{tabular}
\caption{\small Optimized lattice vectors in \AA{} for $\alpha$ phase of SnS, SnSe, GeS, GeSe
along with those of phosphorene (P).
}
\label{vectors}
\end{table}

The calculated lattice parameters are in good agreement with experimental data
for bulk SnS ($\mathbf{a}$=4.33\AA{}, $\mathbf{b}$=3.98\AA{} and $\mathbf{c}$= 11.20\AA{})~\cite{ehm,  wiedemeier}
and bulk SnSe ($\mathbf{a}$=4.44 \AA{}, $\mathbf{b}$=4.15 \AA{}, and $\mathbf{c}$=11.50).\cite{lefebvre}
Our results are also in agreement with previous theoretical studies of SnS~\cite{sebahaddin, vidal, tritsaris}. 
It is interesting to note that the lattice parameters show little variation amongst the four compounds,
differing less than 7\%. 
This is due to the similar electronegativity of Se and S, and of Ge and Sn, and consequently to the similar bond strengths.
Thus, the lattice parameter trend is mostly dominated by the ionic radius of the constituents, 
the most compact structure being that of GeS.
The lattice parameters of the sulphides remain nearly unchanged for different number of layers, while those of the selenides show variations of $\sim \pm$~2\%.

\subsection{Electronic Properties of monolayer, bilayer and bulk models.}

A tunable  bandgap energy within the visible range, is
one of the most interesting properties of the group-IV monochalcogenides.
The calculated values for the energy gaps ($E_g$) are summarized in Table~\ref{gap}.
These were obtained by calculating the bandstructures using the HSE functional along the high-symmetry paths
of the Brillouin zone (BZ) (Fig.\ref{unitcell-bz}-d).
The HSE exchange-correlation functional opens the gap,
comparing to the PBE bandstructure, while the band dispersion remains nearly unchanged

\begin{figure*}[!htb]
    \centerline{
    \includegraphics[scale=1.60]{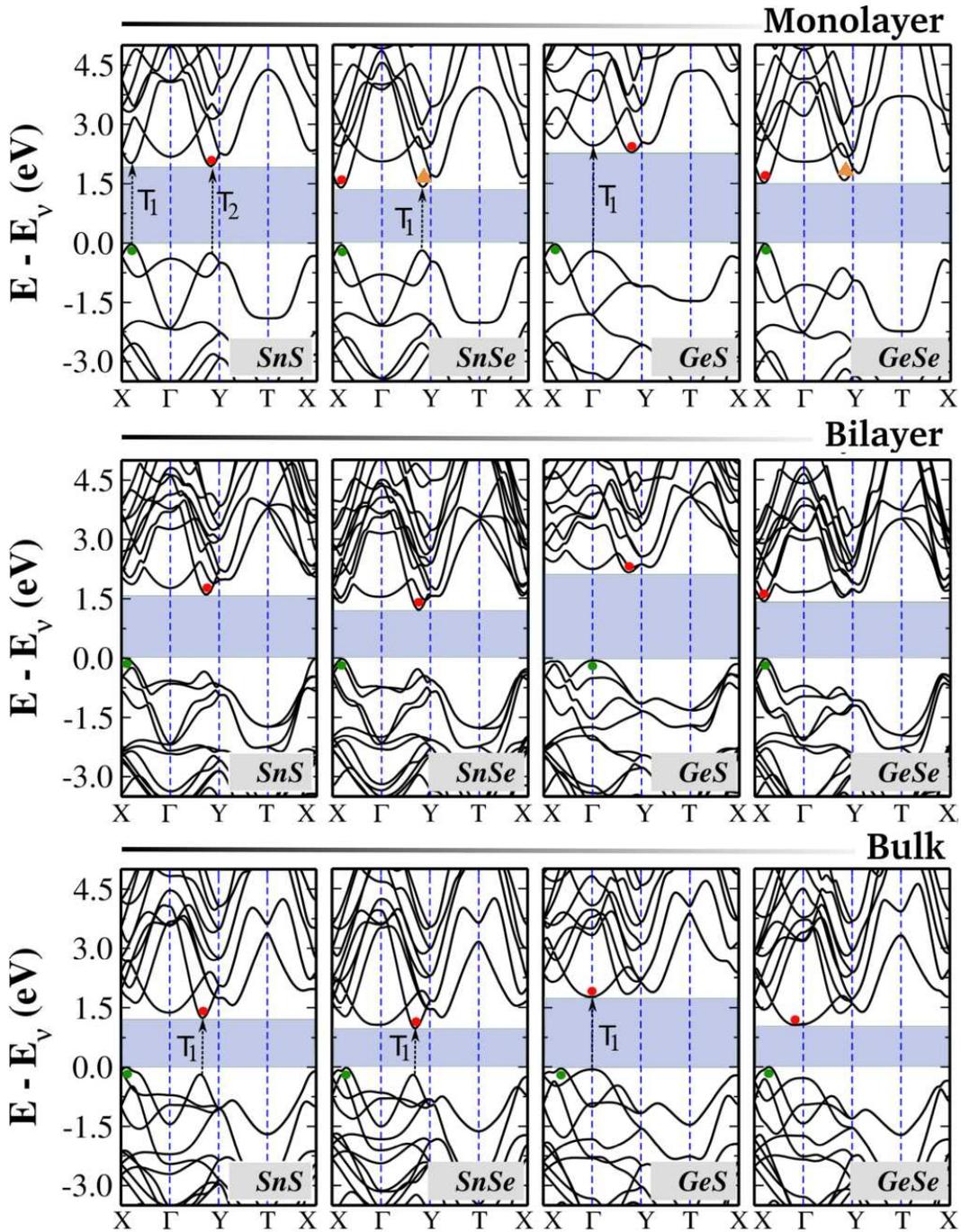}}
   \caption{\small (Color online) Electronic band structures for monolayer, bilayer and bulk group-IV monochalcogenides 
                    calculated using the HSE hybrid functional. The VBM and CBM are highlighted by full circles. Dashed 
                    black arrows indicate possible direct transitions ($T_1$ and $T_2$) to points very close in energy 
                    to the VBM and CBM. Triangles indicate the position of the CBM when spin-orbit coupling effects are considered.}
   \label{bands.HSE}
 \end{figure*}

Some aspects of the bandstructure are common to all systems studied.
The dispersion of the bands nearest to the gap is nearly the same along the $\Gamma$-X and
$\Gamma$-Y directions,
despite the striking difference between the structure along those two crystallographic directions.
This is probably due to the Sn-$5s$ (Ge-$4s$) character of those bands.
Additionally, in most cases there are  multiple valence band and conduction band valleys,
and most of the compounds have indirect gap, except for monolayer SnSe, bulk GeS and monolayer and 
bilayer GeSe, as shown by our calculated bandstructures in Fig.~\ref{bands.HSE}. This is different 
from black phosphorus and phosphorene, which has a well defined direct or nearly-direct gap.
Finally, as expected, $E_g$ decreases with increasing number of layers.

\begin{table}[htb]
\small
\centering
\renewcommand{\arraystretch}{1.8}
\begin{tabular}{m{1.0cm} m{1.0cm} m{1.9cm} m{1.9cm} m{1.9cm}}
\hline   \hline
                      &      &   Monolayer     &    Bilayer      &      Bulk        \\ \hline
\multirow{2}{*}{SnS}  & CBM  & (0.00 ; 0.84)   &  (0.00 ; 0.74)  &  (0.00 ; 0.70)   \\
                      & VBM  & (0.22 ; 0.00)   &  (1.00 ; 0.00)  &  (0.12 ; 0.00)   \\ \hline
\multirow{2}{*}{SnSe} & CBM  & (0.20 ; 0.00)   &  (0.00 ; 0.78)  &  (0.00 ; 0.72)   \\
                      & VBM  & (0.20 ; 0.00)   &  (0.20 ; 0.00)  &  (0.30 ; 0.00)   \\ \hline
\multirow{2}{*}{GeS}  & CBM  & (0.00 ; 0.82)   &  (0.00 ; 0.76)  &  (0.00 ; 0.00)   \\
                      & VBM  & (0.26 ; 0.00)   &  (1.00 ; 0.00)  &  (0.40 ; 0.00)   \\ \hline
\multirow{2}{*}{GeSe} & CBM  & (0.20 ; 0.00)   &  (0.20 ; 0.00)  &  (0.80 ; 0.00)   \\       
                      & VBM  & (0.20 ; 0.00)   &  (0.20 ; 0.00)  &  (0.30 ; 0.00)   \\ \hline  \hline
 \end{tabular}
\caption{\small (Color online) ($\Delta_{k_x}$ ; $\Delta_{k_y}$): Positions of the VBM and CBM along the $\Gamma$-X and $\Gamma$-Y lines in the BZ. 
                The values are given in units of 2$\pi$/$|\mathbf{a}|$ and 2$\pi$/$|\mathbf{b}|$ for \^k$_x$ and \^k$_y$ directions, 
                respectively.}
\label{delta_k}
\end{table}

In the following, we examine in detail the bandstructure of each one of the compounds.
The energy gaps given were obtained with the HSE functional, unless otherwise stated,
and without considering spin-orbit coupling.
Last, we will consider the spin-orbit splitting.

\begin{table*}[!htb]
\small
\centering
\renewcommand{\arraystretch}{1.4}
\begin{tabular}{m{1.2cm} m{1.1cm} m{1.1cm} m{1.8cm} m{1.1cm} m{1.1cm} m{1.8cm} m{1.1cm} m{1.1cm} m{1.1cm} m{2.8cm}}
\hline  \hline  
         &       \multicolumn{3}{ c }{Monolayer}       &\multicolumn{3}{ c }{Bilayer}     &\multicolumn{4}{ c}{Bulk}                                    \\ \hline
&{\scriptsize GGA QE} &{\scriptsize GGA VASP} &{\scriptsize HSE \hspace{0.5 cm} VASP} &{\scriptsize GGA QE}&{\scriptsize GGA VASP} &{\scriptsize HSE \hspace{0.5 cm} VASP} &{\scriptsize GGA QE}&{\scriptsize GGA VASP} &{\scriptsize HSE VASP} & \\ \hline
SnS      & 1.40             & 1.38             &  1.96            &  1.14           &   1.12      &   1.60   &  0.83   &  0.82     &  1.24   & 1.20-1.37~\cite{brad}       \\
SnSe     & 1.01*            & 0.96*            &  1.44*           &  0.79           &   0.76      &   1.20   &  0.55   &  0.54     &  1.00   & 0.898~\cite{prb-41}, 0.95~\cite{crys-15-10278} \\          
GeS      & 1.69             & 1.65             &  2.32            &  1.55           &   1.55      &   2.20   &  1.24   &  1.22     &  1.81   & 1.70-1.96~\cite{brad}       \\
GeSe     & 1.14*            & 1.18*            &  1.54*           &  1.02*          &   0.98*     &   1.45*  &  0.59   &  0.57     &  1.07   & 1.14~\cite{dimitri}         \\ \hline
P        & 0.90             &   -              &  1.66            &  0.55           &     -       &   1.30   &  0.07   &    -      &  0.39   & 0.33~\cite{prb-92, jap-34-1853, jpsj-52-2148} \\ \hline \hline
 \end{tabular}
\caption{\small Gap energies ($E_{g}$) for monolayer, bilayer and bulk monochalcogenides and phosphorene from GGA and hybrid functional calculations. 
                We compare results from Quantum Espresso (QE) and VASP codes for the GGA approach. The star (*) indicates direct band gaps. Experimental 
                values from previous works are also shown. All values are given in eV.}
\label{gap}
\end{table*}
 
\paragraph{\rm SnS} has indirect gap independently of the number of layers.
The indirect bandgaps calculated with HSE are $E_g$~=~1.96, 1.60 and 1.24~eV for monolayer, 
bilayer and bulk SnS, respectively. 
The results for bulk agree very well with the experimental measured gap energies compiled in Ref.~\cite{brad}, 
which presents bandgap energies obtained from optical absorption measurements and extrapolated to zero temperature around 1.2-1.3~eV, agreeing very well with 
our calculated values.
Another previous study\cite{burton-cm} found a bandgap of 1.11~eV for bulk using HSE06, 
but different from our study, they used the experimental lattice parameters as input.
The excellent agreement between the calculated HSE bandgap and the experimental energies
is unexpected, since the gap measured in optical experiments differs from the 
conduction gap by the exciton binding energy, which can be of the order of hundreds of meV in 
two-dimensional materials.\cite{rodin}
However, previous theoretical studies using the GW method~\cite{huser-gw, hedin-gw}, 
obtained $E_g$~=~2.57, 1.57 and 1.07-1.26~eV for monolayer, 
bilayer and bulk SnS, respectively.~\cite{tritsaris,brad}.
This variation in the GW gaps may be due to the difficulty in treating the screening in 2D materials.\cite{huser} 
The agreement is very good except for monolayer, and this difference may be due to several factors. 
One of them is the presence of shallow core $d$ bands in SnS, as discussed in Ref.~\cite{burton-apl}.

We now turn to the details of the bandstructure.
In all cases (monolayer, bilayer and bulk), the valence band maxima (VBM) and conduction band minima (CBM) are located along the $\Gamma$-X and $\Gamma$-Y lines.
In the monolayer, there are other competing local CBM and VBM, very close in energy to the band edges. By considering this, in addition to the 
indirect band gap observed for the monolayer, two direct gaps higher in energy by 75~meV (represented by transition T$_1$ in Fig.~\ref{bands.HSE}), 
and 0.22~eV (T$_2$) are also identified along $\Gamma$-X and $\Gamma$-Y lines, respectively. 
For the bulk, we also observe a competing point along the $\Gamma$-Y line, defining a direct gap of 1.40~eV. 
The situation is similar for the other compounds studied.

\paragraph{\rm SnSe}
For monolayer SnSe, the direct gap of 1.44~eV is calculated along the $\Gamma$-X line.
A second maxima at 0.16~eV above VBM is obtained in the $\Gamma$-Y direction, 
defining an additional direct transition at 1.60~eV (T$_1$). The bilayer bandstructure shows some differences. The CBM is now in the $\Gamma$-Y direction, 
and the material is characterized by an indirect gap of 1.20~eV, with the VBM located in the $\Gamma$-X direction. 

For the bulk model, the CBM is also along the $\Gamma$-Y direction, and an indirect gap of 1.00~eV is defined with the 
VBM in the $\Gamma$-X direction. In this case, the second point nearest in energy to the VBM, lower in energy by 0.17~eV, is 
located along $\Gamma$-Y, and defines a direct gap of 1.17~eV (T$_1$ in Fig.~\ref{bands.HSE}). Both indirect and 
direct calculated gaps agree very well with the experimentally estimated values of 0.95 and 1.15~eV from Ref.~\cite{crys-15-10278}. 

\begin{table}[htb]
\small
\centering
\renewcommand{\arraystretch}{1.4}
\begin{tabular}{m{1.2cm} m{1cm} m{1cm} m{1cm} m{1cm} m{1cm} }
\hline   \hline
           &\multicolumn{2}{ c }{Monolayer}  &{Bulk}     \\ \hline
           &    $T_1$       &        $T_2$   &     $T_1$       \\ \hline
SnS        &    2.03        &    2.18        &     1.40        \\
SnSe       &    1.60        &     -          &     1.17        \\          
GeS        &    2.62        &     -          &     1.83        \\
GeSe       &     -          &     -          &      -          \\ \hline  \hline

 \end{tabular}
\caption{\small Smallest direct gaps obtained from the HSE calculations. The respective transitions are indicated by black arrows in 
                Fig.~\ref{bands.HSE}. The values are given in eV.}
\label{direct-gaps}
\end{table}

\paragraph{\rm GeS}
Monolayer GeS has an indirect band gap of 2.32~eV. This is defined by the CMB and VBM along the $\Gamma$-Y and the $\Gamma$-X lines. 
The lowest energy direct transition is at the $\Gamma$ point (T$_1$ in Fig.~\ref{bands.HSE}). 
This direct gap is only 0.3~eV higher in energy than the indirect gap.
In bilayer GeS, the gap is also indirect and 0.12~eV lower in energy than that of monolayer. 
The CBM in this case is still along $\Gamma$-Y, 
but the VBM is  at the $\Gamma$ point. A second VB maximum at the X point is almost degenerate in energy with the VBM at $\Gamma$.

Bulk GeS has an indirect gap of 1.81~eV. The direct gap at the $\Gamma$ point is only 20~meV higher in energy than the  indirect one.
Our results are in excellent agreement with the  bulk GeS gap energies extrapolated to zero temperature given in Ref.~\cite{brad},
which range from 1.70 to 1.96~eV for different experiments.

\paragraph{\rm GeSe}
For GeSe monolayer, our calculations produce a direct gap $E_g$=1.54~eV along the $\Gamma$-X line. For the bilayer structure, 
a direct gap of 1.45~eV is found in the $\Gamma$-X direction, near the X point. 

Similar to  mono- and bilayer GeSe, the electronic structure for the bulk also shows the VBM and CBM in 
the $\Gamma$-X direction. However, the calculated gap of 1.07~eV is indirect in this case, as the CMB is located closer to the $\Gamma$ point. 
In this case our calculations are also in very good agreement with experimental results, where optical measurements have 
indicated bulk GeSe as an indirect band gap semiconductor with $E_g$=1.14~eV~\cite{dimitri}.

\subsubsection{Spin-orbit splitting}

In the case of monolayer, the spin-orbit coupling gives rise to a splitting of the bands.
This is different from monolayer phosphorene, where the inversion symmetry, together with 
time reversal symmetry, requires that the bands for the two spins are degenerate. No spin-orbit 
splitting is expected in bulk and even-numbered layer group-IV monochalcogenides for the same reason.

\begin{figure}[!htb]
    \centerline{
    \includegraphics[scale=0.30]{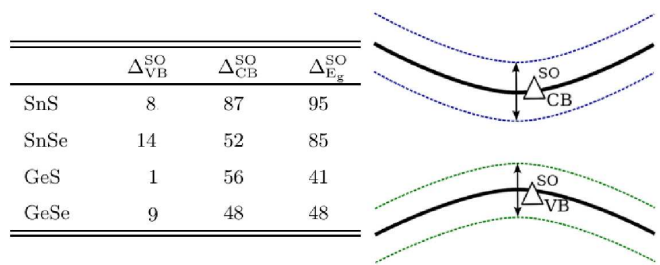}}
   \caption{\small (Color online) Calculated SO-splittings of the CB and VB for the valleys along $\Gamma$-Y in meV. The schematic picture shows in continuous black lines 
                   the bands without spin-orbit effects, which are taken into account in a fully relativistic calculation. The lifted conduction 
                   and valence bands are shown in dashed lines. The lowering in E$_g$ is given by $\Delta_{\rm E_g}^{\rm SO}$.
                   }
   \label{SO.delta}
\end{figure}

\begin{figure}[!htb]
   \centerline{
   \includegraphics[scale=0.30]{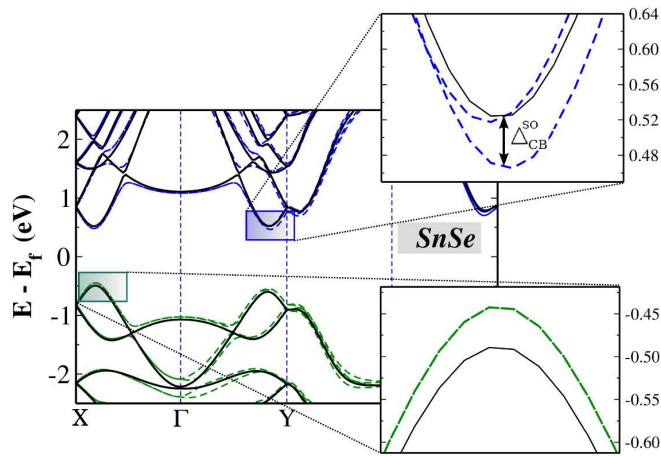}}
  \caption{\small (Color online) Electronic band structures for SnSe monolayer with (green and blue dashed lines) and without (continuous black lines) 
                  spin orbit coupling effect.}
  \label{snse.SO}
\end{figure}

In order to quantify the effect of SO coupling, a fully relativistic calculation based on GGA was performed for the monolayer 
of all compounds. 
Fig.~\ref{snse.SO} shows the calculated electronic bands for monolayer SnSe with and without spin-orbit coupling. 
The results for the other materials, given in SM, are very similar. So, for sake of simplicity, we discuss only the SnSe case. 

The spin degeneracy of the bands is lifted in all BZ except along $\Gamma$-X, the direction along which the $C_2$ rotational symmetry 
and the $xz$ mirror symmetry are preserved. Otherwise the SO-coupling changes little the shape of the bands, except for band crossings 
avoided in the relativistic result.
However, the absolute minimum of the conduction band is along $\Gamma$-Y for all systems.
For SnSe, the conduction band splits by 52~meV near the Y point (at the $\Gamma$-Y line).
This defines a new CBM, since the other maxima at the $\Gamma$-X direction shifts by only 38~meV. 
The same occurs for GeSe, for which a split of 50~meV for the valley near Y is calculated.
The VBM remains along $\Gamma$-X in all cases. The calculated spin orbit splittings are large for the 
conduction bands of all the phosphorene analogues, with the largest calculated value being for the SnS CBM (86~meV), 
as can be seen in the table presented in Fig.~\ref{SO.delta}.
This exceeds the spin-orbit splittings for the conduction bands of some of the most used transition 
metal dichalcogenides, eg. MoS$_2$, MoSe$_2$, WS$_2$ and WSe$_2$, which have spin-orbit splittings between 3 and 30 meV 
at the conduction band valleys.\cite{kosmider}. However, the SO split on the valence band valleys is always
smaller than for TMD, which have splittings of the order of 0.15-0.50~eV.\cite{di-xiao, gui-bin}

\subsection{Optical Properties} 

We now analyze the optical properties of this class of materials,
discussing the influence of anisotropy and the consequences of lower dimensionality.
The complex dielectric function $\epsilon(\omega)$ of the bulk materials
has been measured by electron energy loss spectroscopy (EELS),
optical transmission and reflectance measurements.\cite{makinistian1, wiley, eymard, prb-36-7491} 
It can also be obtained from the optical conductivity $\sigma(\omega)$, which is related to $\epsilon$ by:

\begin{equation}
 \epsilon(\omega) = 1 + \frac{i}{\omega \epsilon_0} \sigma(\omega),  \hspace{0.6 cm} \sigma(\omega) = -i\omega\epsilon_0\left[\epsilon(\omega)-1\right].
 \label{eps-sigma}
\end{equation}
where $\omega$ is the frequency of the incoming electromagnetic wave and $\epsilon_0$ the vacuum permittivity.

The imaginary part of the dielectric tensor $\epsilon_{i}(\omega)_{\alpha,\beta}$ for the bulk system can be obtained from the first-principles bandstructure using
\begin{equation}
\begin{split}
 \epsilon_i(\omega)_{\alpha \beta} = \frac{2 \pi^2 e^2}{m^2V} \sum_{i,f} \int \frac{\mathbf{M}_{\alpha \beta}} 
{\left(E_f({\bf k}) - E_i({\bf k})\right)^2}\\ \times \delta(E_f({\bf k}) - E_i({\bf k}) -\hbar \omega) d^3k
 \label{eps-im}
\end{split} 
\end{equation}
where $E_i({\bf k})$ represents the Kohn-Sham eigenvalues, $\mathbf{M}_{\alpha \beta}$ represents the squared momentum matrix elements, 
$\alpha$ and $\beta$ are the crystal directions, and subscripts $i$ and $f$ correspond to initial and final states, respectively. $V$ and $m$ 
are the cell volume and the electron mass. Note that this includes only direct transitions.

From Eq.~(\ref{eps-im}), the real part of $\epsilon(\omega)$ can be obtained via the Kramers-Kroning relations:

 \begin{equation}
   \begin{split}
 \epsilon_r(\omega)_{\alpha \beta} = 1 + \frac{2}{\pi} \int_0^\infty \frac{\omega' \epsilon_i(\omega')_{\alpha \beta}}{\omega'^2 + \omega^2}d\omega'
   \end{split}
   \label{eps-r}
 \end{equation}

Since the integral over the BZ needs a very large number of points to converge, we use the GGA functional for the calculation of 
$E_i({\bf k}),E_f({\bf k})$, but subsequently apply a rigid shift to correct the bandgap to the value obtained from the HSE calculation.

\begin{figure}[!htb]
    \centerline{
    \includegraphics[scale=0.40]{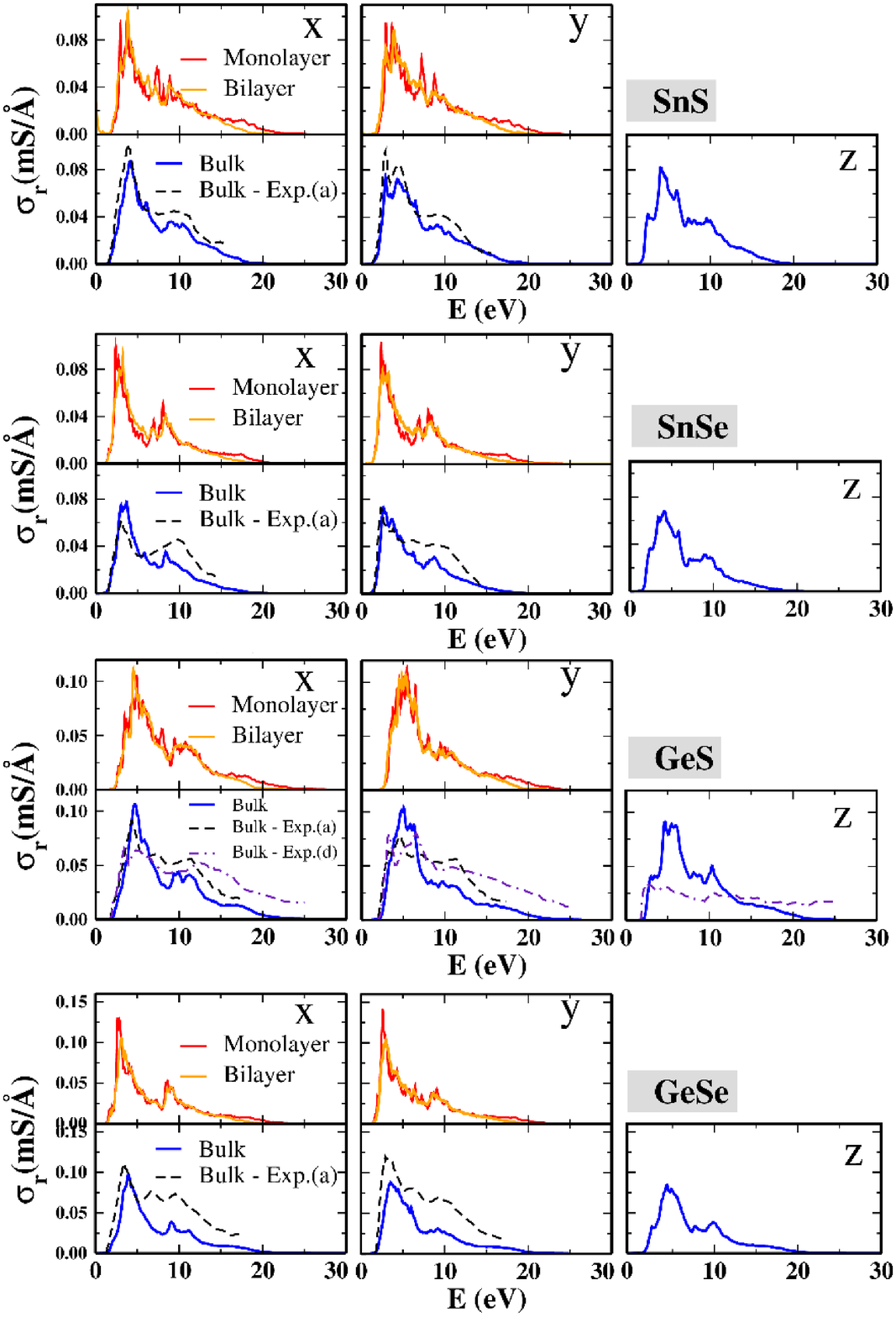}}
    \caption{\small (Color online) Real part of the optical conductivity $\sigma_r(\omega)$ for monolayer (full red lines), 
                    bilayer (full orange lines) and bulk (full blue lines) monochalcogenides. For comparison, experimental 
                    results are also included from previous works. As the experimental works report $\epsilon_i(\omega)$, 
                    we use relations \ref{eps-sigma} to obtain $\sigma_r(\omega)$. Ref(a):~\cite{eymard}; Ref(d):~\cite{wiley}; 
                    Ref(e):~\cite{makinistian1}; Ref(f):~\cite{pss-117}.}
  \label{sigma.mono.bulk}
 \end{figure}

We start by validating the calculation by comparing $\sigma(\omega)$ obtained for bulk with the respective values
extracted from experiment. For all the four monochalcogenides, the agreement is excellent, as shown in Fig.~\ref{sigma.mono.bulk}, 
which compares the calculated real part of the conductivity $\sigma_(\omega)$ with experimental results from previous works.
Both the threshold energy and the bandwidth of the $\sigma_i$ spectrum, which extends to about 20 eV in all cases,
are well reproduced by the calculation.
The low frequency value of the real part, $\sigma_r$, is also in reasonable agreement with experiment.

A striking feature is that the dielectric function is nearly isotropic i.e.,
the three diagonal components $\sigma_{xx}$, $\sigma_{yy}$ and $\sigma_{zz}$ have nearly the
same magnitude and spectral dependence.
This is a direct consequence of the similar band dispersion along the three respective directions of the BZ,
which in turn originates in the structure.
As shown in Fig.~\ref{unitcell-bz}, the Pcmn structure can  be regarded as a distorted cubic NaCl structure,
and the dispersion of the bands near the gap, mostly with $s$ character, is little affected by
the symmetry breaking, specially in plane.
The nearly isotropic behavior of $\epsilon$ is in agreement with previous calculations,\cite{makinistian1,makinistian2}
but at variance with experimental data for GeS\cite{wiley}.
This may be due to the difficulty in performing measurements with the electric field polarization
perpendicular to the layers.
The nearly-isotropic behaviour of the optical response is
in variance with black phosphorus, which has significantly different absorption thresholds 
for the two  in-plane directions.

\begin{figure}[!htb]
    \centerline{
    \includegraphics[scale=0.40]{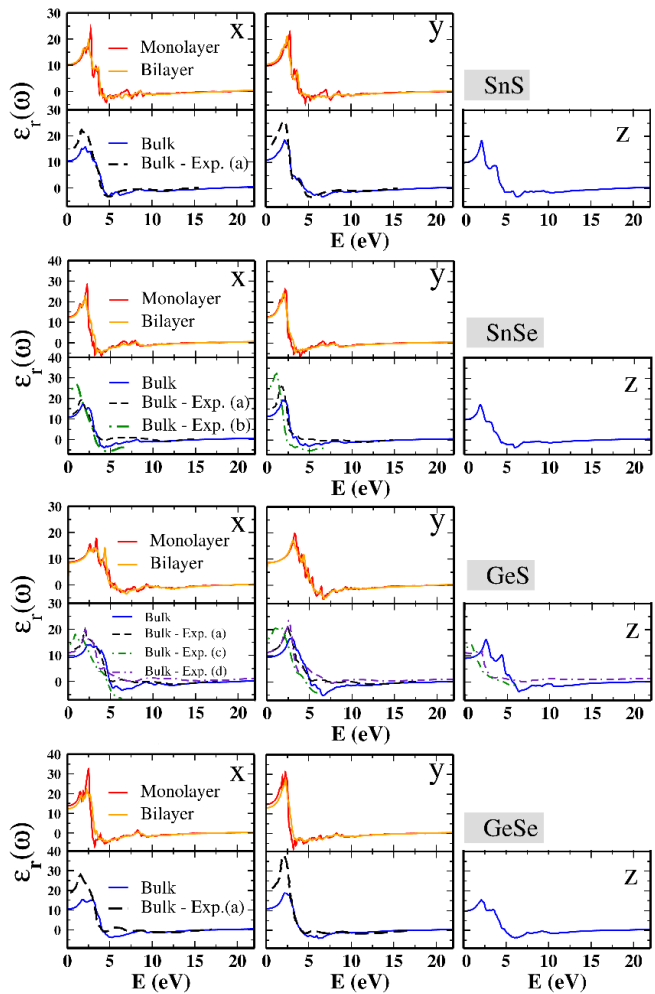}}
   \caption{\small (Color online) Real part of dielectric constants $\epsilon_r(\omega)$ for monolayer (full red lines), 
                    bilayer (full orange lines) and bulk (full blue lines) monochalcogenides. Experimental 
                    results from previous works are also included. Ref(a):~\cite{eymard}; Ref(b):~\cite{prb-36-7491}; 
                    Ref(c):~\cite{prb-31-2180}; Ref(d):~\cite{wiley}.}
   \label{epsilon.mono.bulk}
 \end{figure}

Plasmon energies of these materials can also be estimated from the
first-principles dielectric function, when $Re[\epsilon]=0$. 
A comparison of the calculated real part of the dielectric function $\epsilon_r(\omega)$ with 
experimental results from previous works is presented in Fig.~\ref{epsilon.mono.bulk}.
The plasmon energies are calculated for the intrinsic (insulating) material ie.
inter-band plasmons are considered. The calculated values are in the range 15-20 eV, in excellent 
agreement with experiment,\cite{eymard} except for the sulphides, (Table~\ref{epsr}) for which the plasmon frequency is 
slightly overestimated.
We also note that although experimentally the plasmon frequency is the same for the three dimensions, there is 
a slightly variation in the theoretical results up to the uncertainty of the calculation. Even so, the 
experimental results are very well reproduced.

We now compare the optical response of the monolayer and bilayer material with those of bulk.
For two-dimensional materials, the dielectric constant is not well defined, depending on the interlayer distance $L$ as\cite{cudazzo, berkelbach,huser}
\begin{equation}
\epsilon=1+\frac{4\pi\chi_{2D}}{L}.
\end{equation}

The 2D polarizability $\chi_{2D}$ is constant, and can be obtained from the first-principles calculations,
where $\epsilon$ is calculated from Eq.~\ref{eps-im} by integrating over the whole supercell. 
The interlayer distance $L$ in that case thus corresponds to the supercell
length along the direction perpendicular to the layers.\cite{rodin}
Hence, direct comparison between the 2D and 3D systems is not possible unless $L$ is defined.
Here, in order to obtain effective monolayer and bilayer dielectric constants comparable with bulk, 
thus highlighting the differences originating in the bandstructure and lower dimensionality,
we take $L$ to be the bulk interlayer distance $|${\bf a}$|$/2, thus defining $\epsilon_{\rm eff}=1+(\epsilon-1)\frac{L}{a/2}$.

Comparing the effective optical conductivity for monolayer, bilayer and bulk,
we notice that the spectral dependence and magnitude are very similar,
however, as the number of layers decreases, the peaks get sharper, due to the divergence
of the joint density of states for saddle points of $E_f({\bf k}) - E_i({\bf k})$.\cite{dichalc}
The real part of $\epsilon$ is also nearly unchanged for
effective interlayer spacing equal to that of bulk.
More details of the points of interest in the spectra are given in Table~\ref{epsr}.
           
 \begin{table*}[htb]
\small
\centering
\renewcommand{\arraystretch}{1.4}
 \begin{tabular}{m{1cm} m{0.8cm} m{1.3cm} m{1.3cm} m{0.8cm} m{1.3cm} m{0.8cm} m{0.84cm} m{1.3cm} m{1.3cm} m{0.8cm} m{0.8cm} m{1.2cm} m{1.0cm}}
\hline  \hline  
    &       \multicolumn{5}{ c }{Monolayer}           &       \multicolumn{8}{c}{Bulk}             \\ \hline
    &\multicolumn{2}{c}{$\epsilon_r^0$}&$\epsilon_r^\infty$&\multicolumn{2}{c}{$\omega_p$}  &\multicolumn{3}{c}{$\epsilon_r^0$} &$\epsilon_r^\infty$&\multicolumn{3}{c}{$\omega_p$}&\multicolumn{1}{c}{$\omega_p$-Exp.}\\ \hline
    &  x   &  y    &   x, y            &  x   &   y   &  x  &  y    &  z    & x,y,z          &  x    &   y  &  z    & x,y,z       \\ \hline
SnS &  9.9 & 10.0  &  0.69             & 19.3 & 19.2  &10.30 & 10.94 & 9.90 &   0.72         & 18.1  & 17.8 & 18.0  &  16.22      \\
SnSe& 12.5 & 12.8  &  0.77             & 14.8 & 14.8  &10.90 & 11.50 & 9.90 &   0.78         & 15.7  & 15.7 & 15.9  &  15.44      \\
GeS &  8.7 &  8.6  &  0.63             & 20.4 & 20.4  & 9.20 &  9.45 & 9.10 &   0.64         & 20.2  & 20.1 & 20.2  &  18.21      \\
GeSe& 13.8 & 14.7  &  0.70             & 17.2 & 16.5  &10.60 & 10.96 & 9.80 &   0.72         & 17.7  & 17.2 & 17.8  &  17.30      \\ \hline \hline
 \end{tabular}
\caption{\small Calculated values of the real part of the dielectric constant at $\omega \rightarrow$ 0 ($\epsilon_r^0$) and at high energy limits ($\epsilon_r^\infty$). 
                The plasmon frequencies ($\omega_p$) for {\it x} and {\it y}-directions in monolayer and {\it x}, {\it y} and {\it z}-directions in 
                bulk are also presented. For both monolayer and bulk, $\epsilon_r^\infty$ assume the same values for all directions, while the static dielectric constant $\epsilon_r^0$ 
                and the plasmon frequency $\omega_p$ varies slightly for different directions. Experimental values of $\omega_p$ for the bulk of all compounds from Ref.~\cite{eymard} 
                are also included. All values are given in eV.}
\label{epsr}
\end{table*}

\section{Conclusions} 

We have performed a systematic study of the electronic and optical properties
of the family of phosphorene analogues SnS, SnSe, GeS and GeSe, 
and explored the consequences of lower dimensionality and symmetry breaking.
 
One of the most interesting facets of these materials is that, from bulk down to monolayer, 
they cover a wide range of bandgap energies, from $\sim$~1.0 to 2.3~eV according to hybrid 
functional calculations. The bandgap increases as the number of layers is reduced, 
thus making it is possible to extend the absorption edge up to the green region of the spectrum.

In parallel, this family of materials has the advantage of showing little variation in lattice parameters.
This makes alloying a very promising direction for tuning the  
optical and electronic character of the materials. 

Further, the smallest lattice mismatch between SnS and GeSe compounds, for both lattice parameters, can 
be an indication of these materials as good candidates for formation of hybrid structures.\cite{lu-jiong}
Phosphorene itself has a small lattice mismatch to some of these materials, with which it can be combined.

Additionally, the inversion symmetry breaking in monolayer allows for spin-orbit splitting of the conduction
and valence band valleys along the $\Gamma-Y$ direction. This effect is absent in bulk group-IV monochalcogenides and in phosphorene.
The spin-orbit splitting can be as large as 86 meV for the conduction band minimum of SnS.
The large spin-orbit splitting is likely due to the $s$ character of the VBM and CBM. 

The properties described here open the possibility of using group-IV monochalcogenides
for optoelectronics and spintronics.
Despite, to our knowledge, there being still no experimental reports of isolation of monolayers of 
the monochalcogenides considered here, few-layers isolation by exfoliation has already been achieved \cite{bablu, burton-apl}, 
and it is just a matter of time until control of the layer number becomes possible.

Finally, this illustrates that phosphorene is the parent structure of a whole class of orthorhombic 2D materials,
whose potential is to be revealed once they are isolated in monolayer or few-layer form.

\section*{Acknowledgements}
The authors acknowledge Prof. Antonio Helio Castro Neto for valuable suggestions.
This work was supported by the National Research Foundation, Prime Minister Office, Singapore,
under its Medium Sized Centre Programme and CRP
award ``Novel 2D materials with tailored properties: beyond graphene" (Grant number R-144-000-295-281).
The first-principles calculations were carried out on the GRC high-performance computing facilities.


\begin{thebibliography}{100} 
\bibitem{geim} K. S. Novoselov, A. K. Geim, S. V. Morozov, D. Jiang, Y. Zhang, S. V. Dubonos, I. V. Grigorieva, and A. A. Firsov. Science. {\bf 306}, 666 (2004).
 \bibitem{sheneve} Sheneve Z. Butler, et. al. ACS Nano. {\bf 7}, 2898 (2013).
 \bibitem{lifeng} Lifeng Wang, Bin Wu, Jisi Chen, Hongtao Liu, Pingan Hu, and Yunqi Liu. Advanced Materials {\bf 26}, 1559 (2014).
 \bibitem{qing} Qing Hua Wang, Kourosh Kalantar-Zadeh, Andras Kis, and Michael S. Coleman, Jonathan N. Strano. Nat Nano, 7(11):699–712 (2012).
 \bibitem{tomanek} Han Liu, Adam T. Neal, Zhen Zhu, Zhe Luo, Xianfan Xu, David Tom\'anek, and Peide D. Ye, ACS Nano. {\bf 8}, 4033 (2014).
 \bibitem{rodin} A. S. Rodin, A. Carvalho, and A. H. Castro Neto, Phys. Rev. B {\bf 90}, 075429 (2014). 
 \bibitem{wei-APL-104-251915} Qun Wei and Xihong Peng. Appl. Phys. Lett. {\bf 104}, 251915 (2014).
 \bibitem{fei-Nano Lett-14-6393} Ruixiang Fei, Alireza Faghaninia, Ryan Soklaski, Jia-An Yan, Cynthia Lo, and Li Yang. Nano Lett. 14 (11), pp 6393–6399 (2014).
 \bibitem{wiedemeier} H. Wiedemeier and H. G. von Schnering, Z. Kristallogr. {\bf 156}, 143 (1981). 
 \bibitem{ettema} A. R. H. F. Ettema, R. A. de Groot, and C. Haas. Phys. Rev B. {\bf 46}, 12 (1992).
 
 \bibitem{tritsaris} Georgios A. Tritsaris, Brad D. Malone and Efthimios Kaxiras. J. Appl. Phys. {\bf 113}, 233507 (2013).
 \bibitem{green} A. Green, K. Emery, Y. Hishikawa, W. Warta, and E. D. Dunlop, Prog. Photovoltaics. {\bf 20}, 12 (2012).
 \bibitem{repins} Repins, M. A. Contreras, B. Egaas, C. DeHart, J. Scharf, C. L. Perkins, B. To, and R. Noufi, Prog. Photovoltaics. {\bf 16}, 235 (2008).
 \bibitem{vidal} Julien Vidal, Stephan Lany, Mayeul d'Avezac, Alex Zunger, Andriy Zakutayev, Jason Francis, and Janet Tate. J. Appl. Phys. {\bf 100}, 032104 (2012).
 \bibitem{brad} Brad D. Malone and Efthimios Kaxiras. Phys. Rev B. {\bf 87}, 245312 (2013).
 \bibitem{ktr} K.T. Ramakrishna Reddy, N. Koteswara Reddy, R.W. Miles. Sol. Energy Mater. Sol. Cells. {\bf 90}, 3041 (2006).
 \bibitem{burton-apl} Lee A. Burton and Aron Waish. Appl. Phys. Lett. {\bf 102}, 132111 (2013).
 \bibitem{dong-nature-508-373} Li-Dong Zhao, Shih-Han Lo, Yongsheng Zhang, Hui Sun, Gangjian Tan, Ctirad Uher, C. Wolverton, Vinayak P. Dravid and Mercouri G. Kanatzidis, Nature. {\bf 508}, 373–377 (2014). 
 \bibitem{dresselhaus} J. Ribeiro-Soares, R. M. Almeida, L. G. Can\,cado, M. S. Dresselhaus, A. Jorio, arXiv: 1408.6641.
 \bibitem{bablu} Bablu Mukherjee, Yongqing Cai, Hui Ru Tan, Yuan Ping Feng, Eng Soon Tok, and Chorng Haur Sow. ACS Appl. Mater. Interfaces. {\bf 5}, 9594-9604 (2013).
 
 \bibitem{ks} Konh, W.; Sham, L. J. Self-Consistent Equations Including Exchange and Correlation Effects. Phys. Rev. {\bf 140}, A1133-A1138 (1965).
 \bibitem{Giannozzi2009} P. Giannozzi et al., J. Phys.-Cond. Matter. {\bf 21}, 395502 (2009).
 \bibitem{pbe} Perdew, J. P.; Burke, K.; Ernzerhof, M. Generalized Gradient Approximation Made Simple. Phys. Rev. Lett. {\bf 77}, 3865−3868 (1996).
 \bibitem{tm} Troullier, N.; Martins, J. L. Efficient Pseudopotentials for Plane-Wave Calculations. Phys. Rev. B. {\bf 43}, 1993−2006 (1991).
 \bibitem{Monkhorst1968} H. J. Monkhorst and J. D. Pack, Phys. Rev. B. {\bf 13}, 5188 (1976).
 \bibitem{hse06} J. Heyd, G. E. Scuseria, and M. Ernzerhof, J. Chem. Phys. {\bf 124}, 219906 (2006).
 \bibitem{vasp1} G. Kresse and J. Furthmüller. Comput. Mat. Sci. {\bf 6}, 15 (1996).
 \bibitem{vasp2} G. Kresse and J. Furthmüller. Phys. Rev. B. {\bf 54}, 11169 (1996).
 \bibitem{paw} P. E. Blochl. Projector augmented-wave method. Phys. Rev. B. {\bf 50}, 17953 (1994).
 \bibitem{prb-41} M. Parenteau and C. Carlone, Phys. Rev. B {\bf 41}, 5227 (1990). 
 
 \bibitem{prb-31-2180} S. Logothetidis, L. Vina, and M. Cardona. Phys. Rev. B {\bf 31}, 2180 (1985)
 \bibitem{ehm} L. Ehm, K. Knorr, P. Dera, A. Krimmel, P. Bouvier and M. Mezouar. J. Phys.: Condens. Matter. {\bf 16},  3545-3554 (2004).
 \bibitem{lefebvre} I. Lefebvre, M. A. Szymanski, J. Olivier-Fourcade, J. C. Jumas. Phys. Rev. B. {\bf 58}, 4 (1998). 
 \bibitem{sebahaddin} S. Alptekina, M. Durandurdu. Solid State Comm. {\bf 150}, 17-18, 870-874 (2010).
 \bibitem{burton-cm} Lee A. Burton, Diego Colombara, Ruben D. Abellon, Ferdinand C. Grozema, Laurence M. Peter, Tom J. Savenije, Gilles Dennler, and Aron Walsh. Chem. Mater. {\bf 25}, 4908−4916 (2013).
 \bibitem{huser-gw} F. Huser, T. Olsen, and K. S. Thygesen. Phys. Rev. B. {\bf 87}, 235132 (2013). 
 \bibitem{hedin-gw} L. Hedin. Phys. Rev. {\bf 139} (3A), A796-A823 (1965).
 \bibitem{huser} Falco Hüser, Thomas Olsen, and Kristian S. Thygesen, Phys. Rev. B {\bf 88}, 245309 (2013). 
 \bibitem{crys-15-10278} P. A. Fernandes, M. G. Sousa, P. M. P. Salome, J. P. Leitao, and A. F. da Cunha, Crystengcomm {\bf 15}, 10278 (2013). 
 \bibitem{dimitri} Dimitri D. Vaughn, Romesh J. Patel, Michael A. Hickner, and Raymond E. Schaak. J. Am. Chem. Soc. {\bf 132}, 15170 (2010).
 
 \bibitem{prb-92} Keyes, R. Phys. Rev. {\bf 92}, 580 (1953).
 \bibitem{jap-34-1853} Warschauer, D. J. Appl. Phys. {\bf 34}, 1853 (1963).
 \bibitem{jpsj-52-2148} Akahama, Y., Endo, S. and Narita. J. Phys. Soc. Jpn {\bf 52}, 2148 (1983). 
 \bibitem{kosmider} K. Ko\'smider, J. W. Gonz\'alez, and J. Fern\'andez-Rossier, Phys. Rev. B {\bf 88}, 245436 (2013).       
 \bibitem{di-xiao} Di Xiao, Gui-Bin Liu, Wanxiang Feng, Xiaodong Xu, and Wang Yao. Phys. Rev. Lett. {\bf 108}, 196802 (2012).
 \bibitem{gui-bin} Gui-Bin Liu, Wen-Yu Shan, Yugui Yao, Wang Yao, and Di Xiao. Phys. Rev. B. {\bf 88}, 085433 (2013).               
 \bibitem{wiley} J. D. Wiley, W. J. Buckel, W. Braun, G. W. Fehrenbach, F. J.  Himpsel, and E. E. Koch, Phys. Rev. B {\bf 14}, 697 (1976).
 \bibitem{makinistian1} L. Makinistian and E. A. Albanesi, J. Phys.: Condens. Matter {\bf 19} 186211 (2007).
 \bibitem{eymard} R. Eymard and A. Otto, Phys. Rev. B {\bf 16}, 1616 (1977).
 \bibitem{prb-36-7491} S. Logothetidis and H. M. Polatoglou. Phys. Rev. B {\bf 36}, 7491 (1986)
 
 \bibitem{makinistian2} L. Makinistian and E. A. Albanesi, Phys. Rev. B {\bf 74}, 045206 (2006).
 \bibitem{pss-117} G. Valiukon, et. al. Phys. Stat. Sol.(b) {\bf 117}, 81 (1983).
 \bibitem{cudazzo} P. Cudazzo, I. V. Tokatly, and A. Rubio, Phys.Rev.B {\bf 84}, 085406 ( 2011). 
 \bibitem{berkelbach} T. C. Berkelbach, M. S. Hybertsen, and D. R. Reichman, Phys.  Rev. B {\bf 88} , 045318 (2013). 
 \bibitem{dichalc} A. Carvalho, R. M. Ribeiro, and A. H. Castro Neto, Phys. Rev. B. {\bf 88}, 115205 (2013). 
 \bibitem{lu-jiong} Jiong Lu, Lídia C. Gomes, Ricardo W. Nunes, A. H. Castro Neto, and Kian Ping Loh. Nano Lett. {\bf 14} (9), pp 5133–5139 (2014). 
 
 


\end{thebibliography}
\end{document}